# Unified Architecture Metamodel of Information Systems Developed by Generative AI

Oleg Grynets
EPAM Systems
McLean, Virginia, USA
oleg_grynets@epam.com

Vasyl Lyashkevych
EPAM Systems
Lviv, Ukraine
vasyl_lyashkevych@epam.com

*Abstract*—The rapid development of artificial intelligence (AI) and large language models (LLMs) has driven new methods of software development, in which a large portion of code, technical and business documentation (BD) is generated automatically. However, since there is no single architectural framework that can provide consistent, repeatable transformations across different representation layers of information systems, such systems remain fragmented in their system representation. This study explores the problem of creating a unified architecture for LLM-oriented applications based on selected architectural frameworks by Subject Matter Experts (SMEs)—which are the most popular among vendor's projects, existing design principles and architectural patterns which integrates code, technology diagrams, business models and metadata into a single metamodel. A framework structure is proposed that covers some key types of architectural diagrams and supports a closed cycle of transformations, such as: "Code → Technical documentation (TD) → BD → TD → Code". The key architectural diagrams are split equally between main architectural layers: high-layer (business and domain understanding), middle-layer (system architecture) and low-layer (developer-layer architecture). Each architectural layer still contains some abstraction layers which allow it to be more flexible and fit requirements of design principles and architectural patterns. JSON, PlantUML, and Mermaid formats are defined for representing architectural diagrams, which allows for optimizing algorithmic processing using LLM. The conducted experiments demonstrated stable quality of generated documentations and code when using a structured architectural context in the form of architectural diagrams. The results confirm that the proposed unified architecture metamodel can serve as an effective interface between humans and models, improving the accuracy, stability and repeatability of LLM generation. However, the selected set of architectural diagrams should be optimised to avoid redundancy between some diagrams, and some diagrams should be updated to represent extra contextual orchestration. This work demonstrates measurable improvements for a new generation of intelligent tools that automate the SDLC and enable a comprehensive architecture compatible with AI-driven development.

## I. Introduction

The rapid increase in the complexity of software systems, distributed architectures, microservice ecosystems, the emergence of cloud computing and the recent emergence of code generation systems based on LLMs have brought new challenges to software engineering. In this context, architecture diagrams, as a general-purpose tool, are crucial for representing, analyzing, agreeing on and developing system solutions. They can serve not only as a means of visual communication, but also as a formal mechanism for constructing system models, connecting different layers of abstraction between business requirements, technical architecture, implementation and operating environment.

In classic architectural frameworks such as the Open Group Architecture Framework (TOGAF), the Zachman framework and ArchiMate, diagrams play a central role in modelling business processes, application architecture, technical infrastructure, and the interactions between these layers [1]. They provide architects, analysts, managers and developers with a common language, enabling them to create a unified model of an organization and its information systems.

Simultaneously, modern architectural methodologies have emerged, such as the C4 model, Arc42, Domain-Driven Design (DDD) and the Microservices Architecture Guide. This requires more flexible, engineering-oriented development diagrams that are directly related to the system source code and technical implementation [2].

However, despite the popularity of architectural drawings, their role, necessity and adequacy in modern architecture have not been fully studied. Some scientists have pointed out that diagrams have become "a language used by architects to standardize complex structures and facilitate communication between teams" [3]. Other scholars have pointed out that hand-drawn diagrams are often time-consuming and fail to reflect the true state of the system [4]. With the emergence of DevOps, CI/CD and infrastructure as code, there is a need to synchronize architectural models with actual implementations and operating environments. Modern approaches, such as "architecture as code" or "executable architecture," aim to automatically build diagrams based on telemetry, logs, metadata and code analysis, which significantly alters their role in the development process [5].

Furthermore, the current interest in systems using LLM-based code generation has sparked a new wave of research, focusing on how to use architecture diagrams and models to improve code quality, documentation and design processes. The results show that if LLM can access structured architectural artifacts, such as component diagrams, containers, sequences and domain models, it can produce significantly better code [6]. This means that architecture diagrams are not only a means of knowledge transfer between people, but also an important cognitive interface between people and lifecycle management systems. They provide a structure that helps the model understand the purpose, responsibilities, relationships between services and technical constraints of components.

In microservices and distributed systems, the interactions between many services are extremely complex, and diagrams are increasingly being used as a key tool for scaling planning, monitoring, fault tolerance and optimization. For example, Arc42 includes separate views of runtime interactions, deployment, responsibility assignments and dependencies, enabling architects to systematically analyze the quality of the architecture [7]. At the same time, the C4 model provides a simple and universal approach that is easy for developers to understand and offers a clear hierarchy from context to code.

In the context of digital transformation, the interactions between business processes, systems, services, and teams are becoming increasingly complex, which greatly increases the demand for diagrams. For example, ArchiMate is specifically designed to describe the relationship between business and IT, providing a three-tier model structure: business layer, application layer and technology layer [1]. These diagrams can help to build a holistic model of an enterprise digital twin, analyze the impact of changes and simulate the evolution of the architecture.

Meanwhile, modern scientific research shows that traditional diagrams are no longer sufficient to address new challenges. In particular, the increasing prevalence of low-code or codeless platforms, intelligent agents and generation tools has raised the question of how to create a unified diagram that can be understood by both humans and machines. Researchers emphasize the need for a format that can serve as an intermediate representation between code, documents, architecture and business processes [8]. This format can be a machine-readable diagram, such as Mermaid, PlantUML, JSON, Architectural DSL or another, and can be automatically converted.

Therefore, analysis of modern frameworks and research shows that architecture diagrams are a key element in the process of designing, documenting and developing software systems. They provide multi-layered model integrity, allowing the formalization of system structure and behavior as a basis for analyzing and evaluating architectural solutions, and as a language mediator between business, developers, and intelligent systems. In code generation systems, diagrams play a particularly important role because they can significantly improve the accuracy and quality of automated solutions. This highlights the relevance and necessity of further research aimed at creating a single architectural model to optimize interpretation for both humans and machines.

The scientific problem is that there is currently no way to formally and explicitly represent the knowledge about AI/LLM-based systems or generated by Generative AI (GAI) by means of multi-agent or Agentic AI approaches. There are a lot of challenges in mapping between codebases, TD, business models, and architectural descriptions to be implemented automatically.

In our practice, no widely adopted machine-interpretable framework optimized for LLM transformations that explicitly integrates and formally links the following elements into a single coherent model: a semantic code model; a technical architecture model; business models and business requirements; LLM-based generation mechanisms; code-driven architecture reverse engineering processes; and transformation quality metrics. This makes it impossible to establish a coherent, automated and repeatable software evolution cycle, especially in large projects and mission-critical systems. Therefore, it is essential to be able to create a formal, unified architecture and knowledge representation model for code generation-based systems that enables explicit, repeatable and automated conversion between code and TD, use LLMs to build BD and architectural models, and define metrics and methods for evaluating the quality of such transformations.

Despite extensive research on enterprise architecture, model-driven development, and architecture documentation frameworks, a fundamental gap remains in the explicit unification of business intent, architectural structure, runtime topology, and implementation artifacts into a machine-interpretable metamodel optimized for LLM-driven transformations. Existing methods either focus on business-layer modeling without direct implementation traceability such as TOGAF and ArchiMate or emphasize system structure while neglecting the preservation of semantic intent during iterative generation such as UML-based MDE pipelines. Furthermore, recent research on the development of LLMs has highlighted the lack of structured intermediate representations that can constrain generative models and prevent structural inconsistencies or semantic deviations.

Thus, the purpose of this scientific study is to test the assumption through a controlled comparative evaluation of the transformation process with and without structured architectural mediation.

*Research Assumption:* We assume that, compared to text-centric approaches, integrating a unified multi-layered architectural metamodel as an intermediate representation in a bidirectional LLM-based software transformation cycle significantly improves the quality of code, TD and BD generation. The assumption is being achieved through measurable architectural quality metrics.

## II. RELATED WORKS

Model-driven approaches have long explored the transformation from model to code, particularly through UML-based code generation. However, the systematic evidence suggests recurring limitations in behavioral integrity, traceability and tool fragmentation that restrict end-to-end semantic preservation from design artifacts to implementation [9]. The systematic review [9] of automatic code generation from UML diagrams investigated over 40 methods, tools and frameworks. It has analyzed which UML types were used, which code artifacts such as diagrams, frameworks and partial implementations were created, and recurring gaps such as limited behavior coverage, weak traceability and fragmented tools. This comment reinforces the idea that while there is a process from modeling to code, it often fails to preserve the full semantics and business intent.

A complementary approach [10] focuses on reference architecture and metamodel-driven conformance, where the reference software architecture is formalized using a metamodel. Also, it is being created using a domain-oriented modeling language and converted into skeletal code that conforms to the architecture, reducing erosion and improving repeatability. The work formalizes the Reference Software Architecture (RSA) using a metamodel, defines a subject-oriented graphical modeling language, and generates skeletal code consistent with the RSA [10]. The key idea is to ensure architectural compliance and reduce erosion through model-driven transformation, proving that the metamodel can repeatedly couple architecture to implementation.

Recent research [11] has extended the process in the opposite direction, from code to architecture, by combining reverse engineering with LLM to generate software architecture descriptions from source code, thereby solving the common problem of missing or outdated architecture documentation. This paper proposes a method for semi-automatically generating Software Architecture Descriptions (SADs) from source code by combining reverse engineering and LLMs. The motivation is that SADs are often missing or outdated, leading to increased adoption costs for new users and architectural shortcomings. This approach [11] aims to restore the architectural look and present it in a form that is easy to maintain, thereby improving communication with stakeholders and design justification.

Meanwhile, a systematic review [12] of LLM in software architecture shows that the application of LLM is rapidly growing. Also, it highlights the heterogeneity of cues and assessments, prompting the development of standardized artifacts and protocols. The review summarizes how LLM is used for software architecture tasks such as pattern detection, solution classification or requirements-based architecture design. It reports on the variety of models and stimulation strategies (zero sample or small sample, CoT, RAG) and describes evaluation styles such as experimental, benchmark or case study. The review highlights a lack of consistent evidence on "how well they work", which motivates the usage of standardized artefacts and repeatable assessment protocols.

Beyond functionality, many sources point to a critical uncertainty management issue in LLM software: scaling from prototype to production requires architecture-layer control (validation, governance, drift handling, end-to-end orchestration), not just best guesses [13]. The analysis [13] shows that while LLM systems "look simple" at the prototype stage, they require a system-layer architecture to manage uncertainty, drift, validation and compliance at scale. It emphasizes an integrated workflow mindset, such as "hints → search → orchestration → validation → feedback", and is wary of the "additional tools" thinking pattern.

Empirical evidence from real-world programming environments suggests that users often view LLMs as a hybrid of code generators and information retrieval tools. Verification practices may fail to detect latent errors, which further underscores the necessity of fundamental, verifiable intermediate representations of the architecture [14]. The empirical research [14] shows that scientists frequently use LLM for code generation as an information retrieval tool when reviewing unfamiliar libraries or languages, and also report their verification strategies and potential vulnerabilities.

Formal research [15] shows that learning architectural rules from implementation facts benefits from semantically structured representations: inductive learners can robustly infer constraints in controlled examples, while LLMs can become fragile in confused predicate or negation examples, highlighting the importance of explicit metamodels and machine-interpretable architectural knowledge. This study compares inductive rule learning (IRL) tools with LLM for deriving formalized software architecture rules from implementation facts represented as knowledge bases or metamodels. The results show that IRL can reliably recover the target assumption in relevant paradigms, while LLM's performance is unstable, especially when predicates are confused or negative examples cause training difficulties. This confirms the necessity of constructing structured, semantically relevant architectural representations.

The scientific community also views these issues as ongoing challenges to structured knowledge in terms of coherence, uncertainty, and tool support, providing a theoretical framework for "architecture as bounded knowledge" in the generative SDLC cycle. The work [16] explored the current and future challenges in the field of knowledge representation and reasoning, including reasoning under uncertainty, hybrid reasoning, interpretability and tool support. It provides a theoretical framework for considering architectural knowledge as a structured, bounded representation and highlights the ongoing difficulties of maintaining consistency and handling uncertainty, core issues that also exist in LLM-driven architecture-to-code pipelines.

The chosen design principle determines not only the structural scheme but also the optimal format of the architecture diagram, which is a crucial component for communication, analysis, and the SDLC. However, the automatic generation of such architecture diagrams and their conformity with design principles remains a significant challenge for research and practice [17-19]. Paper [17] studied the role of sketches and diagrams in development practice and showed that diagrams support architectural understanding, communication and design thinking in the design process. A more systematic literature review presented in [18] showed how the choice of architectural mode affects the choice of document models and diagrams and why formal architectural knowledge is important. In [19] a method for the automatic reconstruction of architecture diagrams from code artifacts was proposed, highlighting the importance of consistency between the model and the architectural pattern – automation presents some challenges.

The [20] emphasizes that architectural patterns define a vocabulary of components and relationships that should subsequently be represented in the form of diagrams. In other words, the form and content of the diagrams depend on the chosen pattern, as this determines the focus on specific aspects of the architecture such as components, modules, relationships, data flow, etc. – different patterns require different visualization methods. For example in [20]:

- MVC/Layered Architecture – it is recommended to show the hierarchical structure and dependencies between layers in this architecture such as UML components, C4 components/containers, etc.
- Event-Driven Architecture/CQRS Architecture – Visualizing event flows and asynchronous interactions, such as sequence, activity, event flowcharts, is crucial.
- Microservices Architecture – Deployment diagrams and context/container diagrams (C4) are very useful.

This confirms the necessity of maintaining consistency between graphical representations and pattern semantics. Architectural patterns are widely used in practice, which should be reflected in diagrams [20]. A systematic review [21] shows that different visualizations are chosen depending on the purpose of representing the architecture and its patterns and for representing architectural elements, and that different visualization techniques are classified by type.

Diagrams play an important role in explaining architectures, as they help to reveal structures, solutions, and patterns. Despite their importance, questions remain about their quality and standardization [22]. UML and other diagrams are used to evaluate architectural models and their quality, which considers some appropriate requirements for visual representations [23].

In the studies examined, Specification-Driven Development (SDD) consistently positions architectural sketches as the core bridge between abstract specifications and executable software. Literature indicates that specifications themselves, whether formal or semi-formal, are often too abstract to be directly implemented, while architecture provides the necessary structural decomposition and perspective separation, thereby operationalizing the specifications [24-25].

The study [26] introduces the Software Functional State (SFS) concept, which formalizes software systems as dynamic state-driven entities throughout the SDLC. Unlike purely structural architecture approaches, SFS provides measurable indicators of functional integrity, semantic stability, and SDLC evolution. Integrating SFS with architecture-based intermediate representations may enable state-aware, transformation-stable generative software engineering, where architectural constraints and functional state metrics jointly regulate AI-driven development cycles.

This perspective is directly relevant to current research on unified architectural metamodels and intermediate representations. While this work focuses on structural and semantic invariance in transformation loops "Code↔TD↔BD↔Code", the SFS method extends this perspective by introducing a state-based analysis layer that can assess the stability and robustness of such transformations over time.

Most scientific articles, generally, consider architectural sketches to be indispensable in the software development process: they translate intent into structure, enable traceability and verification, and lay the foundation for the systematic (and increasingly automated) derivation of software from specifications. The architecture diagrams reduce ambiguity, improve traceability and enable incremental refinement from requirements through design to code. An architectural view serves both as a cognitive framework for developers and as a technical framework for automation, allowing tools to verify consistency and manage transitions.

Literature overview indicates that significant progress has been made in software architecture and automation across various fields. Model-based design (MBD) aims to translate models into code, but often faces semantic loss and tool fragmentation. Reference architectures improve architectural consistency but are typically still domain-specific. Reverse engineering combined with LLM methods shows potential in extracting architectural knowledge from source code, but cannot guarantee the formal invariance required for transformation stability.

A recent systematic review of the application of logical reasoning models in software architecture highlights the diversity of evaluation strategies and the lack of standardized intermediate representations suitable for generative reasoning. Empirical studies show that artifacts generated by LLMs based solely on literal clues can exhibit structural inconsistencies, inconsistent dependency directions, and insufficient traceability.

It is noteworthy that while architectural patterns and design principles define structural forces and constraints, few empirical studies have demonstrated how structured architectural diagrams, formalized as a unified metamodel, affect LLM-based regeneration cycles.

Therefore, existing research focuses only on isolated aspects of architecture-to-code or code-to-architecture transformations, lacking an integrated, multi-layered metamodel specifically optimized for bidirectional, LLM-constrained transformations between the business, structural, behavioral, and runtime layers.

This gap immediately led us to propose the assumption of this study: the measurable impact of a unified architectural metamodel on semantic robustness and regenerative stability throughout the entire AI-based SDLC.

### III. METHODOLOGY

#### A. Research Design

This study uses a controlled comparative experimental design aimed at evaluating the impact of structured architectural mediation on software regeneration processes guided by LLM.

Two transformation pipelines were identified:

- Process A (Text-centric pipeline). A transformation cycle based solely on text artifacts.
- Process B (Architecture-mediated pipeline). A transformation cycle that includes machine-readable architectural diagrams as structured intermediate representations.

Both pipelines implement the same closed-loop transformation: "Code → TD → BD → TD → Regenerated Code". Therefore, the independent variable is:

- The architecture diagram exists as a structured intermediate constraint.
- The dependent variable is determined through architecture quality metrics.

#### B. Theoretical methods

Applying system analysis methods to study modern architectural frameworks, such as TOGAF, ArchiMate, C4 model, Arc42, UML, DDD and CNCF to identify their structural components and attributes related to code generation and documentation tasks. Formal modeling techniques are used to create a single architectural metamodel that integrates already existing architectural diagrams into a unified representation. The formalization process is based on entity definitions, relation definitions, semantic constraint definitions and mapping rules between different layers of abstraction.

This paper considers information theory and semantic analysis to identify metrics for information content in architectural models and their impact on the quality of LLM-based code generation. The metrics used include semantic accuracy, structural consistency and model completeness.

#### C. Applied methods

Prompt engineering, including learning templates, thought processes, and structured prompts to help LLMs create code, documentation and architecture diagrams.

Reverse engineering is used for automatic extraction of the architectural structures from the original code and converting it into one of the required formats such as Mermaid, JSON or PlantUML. Static analysis methods are supplemented by dynamic interactive tracing. This allows us to explore the correspondence between code and architecture diagrams.

Software engineering methodologies, such as test-driven development, architecture-driven development and model-driven engineering, are used to build system prototypes and achieve a closed loop: "Code → TD → BD → TD → Code". The architectural diagrams are part of technical and BD as well.

The evaluation comparison, through comparative architecture analysis, of the unified architecture with traditional methods, identified the advantages of the proposed framework.

*D. Evaluation Protocol*

High-quality architectural diagrams are complete, semantically accurate, internally consistent, traceable at different layers of abstraction, machine-readable, and can significantly improve the outcome of LLM-based code generation. To reduce subjectivity and ensure analytical reproducibility, the evaluation indicators were operationalized using measurable indicators.

To achieve a systematic and reproducible evaluation of architecture diagrams, especially in the context of automated and LLM-supported software development, we have introduced a concise yet comprehensive set of quality metrics. These metrics aim to balance analytical rigor with practical applicability, ensuring that architectural artefacts can be evaluated consistently across different systems, domains and architectural frameworks.

*Completeness* measures the extent to which an architecture diagram includes all the relevant system elements and relationships needed to achieve its intended purpose. Incomplete representations introduce ambiguity and limit human understanding and automated processing. Completeness quantifies the structural coverage of system entities represented in the architectural model:

$$C = \frac{|E_{represented}|}{|E_{expected}|}, \quad (1)$$

where: $E_{expected}$ equals to quantity of entities extracted via static analysis of the original codebase and $E_{represented}$ equals the quantity of entities captured in diagrams / metamodel.

*Semantic fidelity* assesses how accurately architectural elements and relationships reflect the actual semantics of the system, including responsibilities, interactions, and domain meaning. High semantic accuracy is crucial for preserving intent during the translation process between business specifications, technical designs, and code. Semantic fidelity measures the preservation of conceptual meaning between stages of transformation:

$$SF = cosine(Embd_{original}, Embd_{regenerated}). \quad (2)$$

Embeddings come from: domain entities, component responsibilities, API contracts. Semantic similarity is computed using vector representations.

*Consistency* assessment assesses internal consistency within a single diagram and across multiple architectural representations. A consistent representation ensures that element definitions are not contradictory or ambiguous when viewed from different layers of abstraction. Consistency $K$ evaluates structural invariance via constraint validation:

$$K = 1 - \frac{V_{violations}}{V_{constraints}}, \quad (3)$$

where: $V_{violations}$ detected architectural rule violations and $V_{constraints}$ total defined constraints.

*Traceability coverage* quantifies the proportion of architectural elements that can be linked across different layers of abstraction, such as from business functions through system components to implementation artefacts. This metric is crucial for supporting specification-driven development and bidirectional conversion. Traceability Coverage $TC$:

$$TC = \frac{|Mapping_{valid}|}{|Mapping_{possible}|}, \quad (4)$$

where: mappings include: "BusinessCapability ↔ Container", "DomainEntity ↔ DataSchema", "Component ↔ CodeModule".

*Machine-readability* refers to the extent to which an architectural diagram is represented in a structured or semi-formal format suitable for automated analysis, verification, and transformation. Without a machine-readable representation, automation and large-scale analysis are not possible. Machine Readability $MR$:

$$MR = \frac{|Artefacts_{parsable}|}{|Artefacts_{total}|}. \quad (5)$$

Parsability determined through deterministic syntax validation.

*LLM Constraint Effectiveness* measures the effectiveness of architecture diagrams in restricting and managing large language models during generation or analysis tasks. This metric reflects the ability of architecture diagrams to reduce ambiguity, avoid artefacts, and improve the structural correctness of LLM results. Constraint effectiveness $LCE$ evaluates reduction of generative structural deviations:

$$LCE = 1 - \frac{D_{drift}}{D_{baseline}}, \quad (6)$$

where: drift defined via: Dependency Graph Edit Distance and Constraint Violation Rate.

*Code coverage by schema* measures the degree to which structural elements in an implementation (such as modules, interfaces, and patterns) can be directly derived from or refactored from the architectural representation. Higher coverage indicates a stronger correspondence between the architecture and the implementation. Patterns detected using static architectural pattern recognition $CPC$:

$$CPC = 1 - \frac{|Patterns_{preserved}|}{|Patterns_{expected}|}, \quad (7)$$

Each system goes through the following stages:

- Extraction of the base architecture;
- Regeneration using process A;
- Regeneration using process B;
- Metrics calculation.

In parallel, the evaluation by SMEs was retained for better evidence. Evaluation of these seven metrics followed a structured ordinal scoring scheme for SMEs: not supported (0), very weak (1), weak (2), good (3), strong (4), excellent (5). The explanation of these metrics is provided in Table 1.

TABLE I. Metrics Measurement Explanation

| Metric | 0 — Not Supported | 1 — Very Weak | 2 — Weak | 3 — Good | 4 — Strong | 5 — Excellent |
|---|---|---|---|---|---|---|
| Completeness | Diagram omits core system elements; architecture unusable for reasoning or generation | Major structural elements missing; relationships largely undefined | Core components present but key dependencies/interactions missing | Most essential elements and relationships represented; minor gaps remain | Nearly all relevant entities, dependencies, and views represented | Full structural coverage aligned with diagram purpose; no meaningful omissions |
| Semantic Fidelity | Diagram contradicts system behavior or responsibilities | Frequent semantic mismatches between diagram and system intent | Partial semantic alignment; ambiguities in roles/responsibilities | General semantic correctness with limited inconsistencies | Strong semantic alignment with business/system intent | Precise preservation of responsibilities, interactions, constraints, and domain meaning |
| Consistency | Contradictory definitions or incompatible views | Multiple internal conflicts across diagram elements | Minor conflicts between relationships or abstractions | Internally coherent with rare inconsistencies | High cross-view coherence; relationships stable | Fully invariant across layers/views; no contradictions |
| Traceability Coverage | No links across abstraction layers | Traceability mostly implicit or ambiguous | Partial mapping between layers (Biz↔System↔Code) | Clear mapping for key elements | Extensive cross-layer mappings with minimal gaps | Complete bidirectional traceability across entities, behaviors, and constraints |
| Machine-Readability | Free-form drawing/text only | Representation difficult to serialize/process | Partially structured format with inconsistencies | Structured notation (PlantUML/Mermaid/JSON) usable with manual correction | Fully structured & parsable representation | Deterministic, loss-minimizing encoding suitable for automation & analysis |
| LLM Constraint Effectiveness | Diagram provides no meaningful constraint or guidance to LLM | Weak guidance; high ambiguity/hallucination risk | Limited constraint effect; LLM frequently misinterprets structure | Diagram improves generation stability in common cases | Diagram reliably constrains structure & dependencies | Diagram acts as a strong generative control layer minimizing drift & hallucination |
| Code Coverage by Schema | Diagram cannot inform code structure | Minimal correspondence with implementation artifacts | Partial mapping to modules/components | Diagram supports skeletal code generation/refactoring | Strong mapping to implementation structure | High-precision derivation of modules, interfaces, schemas, and patterns |

Assessments were performed by one human software architect and two senior engineers who are our SMEs. Also, we have used LLM-based scoring only as an auxiliary analytical reference.

*E. Generation Pipeline Architectures*

To experimentally verify the proposed method, we chose two different transformation pipelines:

- Process A: Text-centric (without architecture diagrams). This pipeline is based solely on text artifacts. Architectural information is conveyed in the form of descriptions, without a formalized structural middleware.
- Process B: Architecture-extended (using architecture diagrams as intermediate representations with some textual information). In this approach, architectural diagrams serve as a formalized intermediate representation that is: human-readable; machine-readable (Mermaid, PlantUML, JSON or DSL) and a stable structural outline for the LLM.

Both processes implemented a closed transformation cycle: "Code → TD → BD → TD → Code".

Execution phase of Process A:

- Code → TD (Fig. 1). Reverse engineer the code (static analysis of dependencies, APIs, modules, and configurations). This generates a written TD.
- TD → BD (Fig. 2). Abstract the technical description to business goals, processes, roles, and requirements.
- BD → TD (Fig. 3). Translate business requirements back into technical specifications. The greatest risk of semantic loss occurs at this stage: written requirements often cannot be clearly mapped to system components.
- TD → Code (Fig. 4). Generate or update code based on the written technical description.

The process is feasible, but its stability largely depends on the quality of the textual description and human verification. It does not include a formal mechanism for

preserving structural invariance. Thus we would have some associated risks:

- semantic discrepancies between hierarchical layers;
- unstable architectural relationships;
- structural "inconsistency";
- poor traceability (difficulty in determining which requirement led to which module).

```
prompt_A_1 = """
You are a senior software architect performing architecture reconstruction.

INPUT:
Source code fragments / repository description:

[INSERT CODE / MODULE DESCRIPTION]

TASK:
Generate formal Technical Documentation (TD) describing the system.

STRICT REQUIREMENTS:

1. Describe ONLY what can be logically inferred from the code.
2. DO NOT invent components, services, layers, or responsibilities.
3. If information is missing, explicitly state: "Not derivable from code".
4. Separate the documentation into sections:

   • System Overview
   • Architectural Style (if derivable)
   • Core Components / Modules
   • Responsibilities of Components
   • Interfaces / APIs
   • Dependencies
   • Data Structures (if visible)
   • Runtime Behavior (ONLY if derivable)
   • Explicit Assumptions
   • Uncertainty / Missing Information

5. Maintain architectural neutrality:
   Avoid guessing patterns (e.g., microservices, CQRS) unless explicitly observable.

6. Use precise technical language.
7. Avoid marketing, speculation, or design recommendations.

OUTPUT FORMAT:
Structured technical document.
"""
```

Fig. 1. Process A: Example of the prompt for TD generation from code.

The execution of Process B assumes the existence of a formally defined unified architecture metamodel and its corresponding set of diagrams.

```
prompt_A_2 = """
You are a business analyst translating Technical Documentation into Business Documentation.

INPUT:
Technical Documentation:

[INSERT TD]

TASK:
Generate Business Documentation (BD).

STRICT REQUIREMENTS:

1. Preserve semantic fidelity with TD.
2. DO NOT introduce features not supported by TD.
3. Translate technical constructs into business meaning:

   • Components → Business Functions
   • APIs → Business Capabilities / Interactions
   • Data Structures → Business Entities (ONLY if justified)

4. Structure BD into:

   • Business Context
   • Business Objectives (ONLY derivable)
   • Business Capabilities
   • Key Business Processes
   • Actors / Roles
   • Value Delivered
   • Business Constraints
   • Explicit Traceability Links to TD
   • Explicit Assumptions
   • Uncertainty / Non-derivable Elements

5. If mapping is ambiguous → explicitly declare uncertainty.
6. Avoid speculative business narratives.
7. Avoid adding domain knowledge not grounded in TD.

OUTPUT:
Formal Business Documentation.
"""
```

Fig. 2. Process A: Example of the prompt for BD generation from TD.

```
prompt_A_3 = """
You are a solution architect converting Business Documentation into Technical Documentation.

INPUT:
Business Documentation:

[INSERT BD]

TASK:
Generate Technical Documentation (TD).

STRICT REQUIREMENTS:

1. Treat BD as authoritative business intent.
2. DO NOT invent technical mechanisms beyond BD scope.
3. Explicitly map:

   • Business Capabilities → System Functions
   • Business Processes → System Interactions
   • Business Entities → Data Structures (if justified)

4. Structure TD into:

   • System Purpose
   • Functional Architecture
   • Logical Components
   • Responsibilities
   • Interfaces / Contracts
   • Data Model (ONLY if derivable)
   • Dependencies
   • Constraints Derived from Business Rules
   • Traceability Matrix (BD ↔ TD)
   • Explicit Assumptions
   • Identified Ambiguities

5. If multiple technical interpretations exist → list alternatives.
6. Avoid premature pattern commitment.
7. Avoid adding hidden infrastructure complexity.

OUTPUT:
Architecturally consistent TD.
"""
```

Fig. 3. Process A: Example of the prompt for TD generation from BD.

Unlike text-based transformation pipelines, the architecture-mediated approach fundamentally relies on structured, machine-interpretable representations that serve as explicit semantic and structural constraints for LLM-based operations.

```
prompt_A_4 = """
You are a senior software engineer generating code strictly from Technical Documentation.

INPUT:
Technical Documentation:

[INSERT TD]

TASK:
Generate implementation artefacts.

STRICT REQUIREMENTS:

1. Follow TD as a specification, not inspiration.
2. DO NOT add features, layers, optimizations, or frameworks not defined.
3. Maintain strict mapping:

   • Components → Modules / Classes
   • Interfaces → Contracts / APIs
   • Data Structures → Schemas / Objects

4. Enforce architectural discipline:

   • Preserve dependency direction
   • Preserve component boundaries
   • Preserve responsibilities

5. If TD lacks implementation detail → use minimal neutral scaffolding.
6. Explicitly document:

   • Generated assumptions
   • Uncertain interpretations
   • Extension points

7. Avoid framework bias unless specified.
8. Avoid performance optimizations unless required.

OUTPUT:
Clean, minimal, architecture-consistent code skeleton.
"""
```

Fig. 4. Process A: Example of the prompt for TD generation from code.

A unified architectural metamodel defines the basic constructs required for deterministic transformations, including entity definitions, relational semantics, abstraction layers, constraint rules, and allowed correspondences between system representations. Without such a metamodel, process B cannot be consistently applied because

LLM-driven transformations require explicit, serialized representations of architectural knowledge.

This dependency arises from the operational characteristics of LLMs. Since LLMs do not have a permanent architectural state, all structural invariants, dependency constraints, and semantic relations must be explicitly provided in the query context. The architectural diagrams derived from the metamodel can serve as a compact encoding of the structure and semantics of the system, thereby ensuring controlled generation and reducing the risk of semantic drift or structural inconsistencies.

Thus, diagrams in Process B not only represent document artifacts, but are also an important component of the generative management interface. The metamodel ensures that semantic consistency, cross-level traceability, and transformation stability of diagram representations are maintained throughout the generation cycle.

For this reason, the following sections present examples of prompts supporting Process B. These prompts explicitly include machine-readable architectural diagrams as structured constraints governing LLM-mediated transformations. Execution phase of Process B:

- Code → TD + Diagrams (Fig. 5). A set of basic diagrams (Context, Container, Component, Data, Integration, Deployment) is formed in parallel with the generation of TD. Diagrams act as the "primary structural constraint layer" of the structure.
- TD + Diagrams → BD + Diagrams (Fig. 6). BD is formed based on business-layer diagrams (Context, Capability, Process). The mapping between the business layer and system elements is preserved.
- BD + Diagrams → TD + Diagrams (Fig. 7). Requirements updates are first reflected in the diagrams, after which the text documentation is synchronized. Thus, the ambiguity of the transformation is reduced.
- TD + Diagrams → Code (Fig. 8). Code generation takes into account structural invariants: containers, components, contracts, data schemas, dependency constraints.

Architectural diagrams provide the role of architectural invariants that control regeneration systems and ensure process reproducibility. Key difference of this Process B - diagrams provide a structural layer of constraints that:

- ensure consistency between layers;
- your traceability;
- reduce the risk of hallucination;
- stabilize two-way transformations.

After completing the cycle for each approach, we compared the results using the evaluation protocol described above.

*F. LLM Control Stack Model*

Although Process B introduces structured architecture diagrams as explicit generation constraints, rapid design remains a key factor for transformation stability and outcome quality. A unified architecture metamodel does not eliminate the need for rapid optimization; rather, it shifts the focus from ambiguity management to constraint-aware orchestration.

Even when using a diagram-based generation approach, the behavior of LLM is very sensitive to the structure of the instructions, the order of the instructions, the contextual emphasis, and the reasoning strategy used. Changes in the representation of the instructions can affect the semantic interpretation, the precedence of constraints, and the preservation of invariants. Therefore, systematic optimization of instructions is an integral part of the architecture-driven SDLC pipeline.

```
prompt_B_1 = """
You are a senior software architect performing model-driven architecture reconstruction.

INPUT:
Source Code / Repository Description:

[INSERT CODE / MODULES / REPO]

TASK:
Generate Technical Documentation (TD) AND machine-readable architectural diagrams.

MANDATORY DIAGRAM SET:

BUSINESS / CONTEXT LAYER:
• Business Context Diagram (C4 C1)
• Business Capability Map (ONLY if derivable)
• Domain Model Diagram (ONLY if derivable)

SYSTEM LAYER:
• System Container Diagram (C4 C2)
• Component Diagram (C4 C3 / UML Component)
• Integration / API Interaction Diagram

IMPLEMENTATION LAYER:
• Class / Module Structure Diagram (UML Class)
• Data Model / Schema Diagram (ERD) (if derivable)

BEHAVIORAL LAYER:
• Sequence / Interaction Diagram (ONLY for derivable flows)
• State Machine Diagram (ONLY if stateful behavior observable)

RUNTIME LAYER (ONLY if derivable):
• Deployment / Infrastructure Diagram
• Runtime Topology / Operational Diagram

STRICT RULES:

1. Diagrams MUST be consistent with TD.
2. DO NOT invent entities not derivable from code.
3. Explicitly declare uncertainties.
4. Preserve dependency direction.
5. Use machine-readable formats:

    • C4 / Structural → PlantUML or Mermaid
    • ERD → Mermaid ER / PlantUML
    • Sequence / State → PlantUML / Mermaid

OUTPUT:

1. Structured Technical Documentation
2. Serialized diagrams grouped by abstraction layer
3. Explicit Assumptions / Uncertainty Section
"""
```

Fig. 5.   Process B: Example of the prompt for TD generation from code.

```
prompt_B_2 = """
You are a business analyst performing architecture-aware business abstraction.

INPUT:

• Technical Documentation
• Architectural Diagrams:

[INSERT TD + DIAGRAMS]

TASK:
Generate Business Documentation (BD) AND business-layer diagrams.

MANDATORY BUSINESS DIAGRAMS:

• Business Context Diagram (refined)
• Business Capability Map
• Business Process Diagram (BPMN)
• DDD Context Map
• Domain Model Diagram (refined semantic version)

STRICT RULES:

1. BD MUST preserve semantic fidelity with TD & diagrams.
2. Business elements MUST map to architectural constructs.
3. DO NOT introduce capabilities unsupported by architecture.
4. Explicitly maintain traceability:

    • Capability ↔ Container
    • Process ↔ Sequence / Interaction
    • Domain Entity ↔ Data Model / Classes

5. Resolve ambiguities explicitly.
6. Preserve system boundaries defined in Context / Container diagrams.

OUTPUT:

1. Formal Business Documentation
2. Business-layer diagrams (machine-readable)
3. Traceability Matrix (BD ↔ Architecture)
4. Explicit Assumptions / Uncertainty
"""
```

Fig. 6.   Process B: Example of the prompt for BD generation from TD.

```
prompt_B_3 = """
You are a solution architect performing constraint-driven technical synthesis.

INPUT:

• Business Documentation
• Business Diagrams:

[INSERT BD + DIAGRAMS]

TASK:
Generate Technical Documentation (TD) AND synchronized architectural diagrams.

MANDATORY DIAGRAMS:

SYSTEM STRUCTURE:
• System Container Diagram (C4 C2)
• Component Diagram (C4 C3)
• Clean / Onion Architecture Diagram
• Integration / API Interaction Diagram

PATTERN LAYER (ONLY if derivable):
• CQRS Diagram
• Event-Driven Diagram

IMPLEMENTATION LAYER:
• Class / Module Structure Diagram
• Data Model / Schema Diagram

BEHAVIORAL LAYER:
• Sequence / Interaction Diagram
• State Machine Diagram (if needed)

RUNTIME (if specified):
• Deployment Diagram
• Runtime Topology Diagram

STRICT RULES:

1. Preserve Business Capability semantics.
2. Enforce architectural invariants.
3. Maintain dependency direction constraints.
4. Multiple interpretations → enumerate alternatives.
5. DO NOT introduce unnecessary infrastructure complexity.

OUTPUT:

1. Technical Documentation
2. Serialized diagrams
3. BD ↔ TD Traceability Matrix
4. Constraint / Invariance Declaration
"""
```

Fig. 7. Process B: Example of the prompt for TD generation from BD.

```
prompt_B_4 = """
You are a senior software engineer performing architecture-constrained code generation.

INPUT:

• Technical Documentation
• Architectural Diagrams:

[INSERT TD + DIAGRAMS]

TASK:
Generate code strictly aligned with architecture.

MANDATORY CONSTRAINT SOURCES:

• System Container Diagram → Services / Repositories
• Component Diagram → Modules / Boundaries
• Clean / Onion Diagram → Dependency Rules
• Integration Diagram → Contracts / Clients
• CQRS Diagram (if present) → Command/Query Separation
• Event-Driven Diagram (if present) → Messaging Logic
• Class Diagram → Code Skeleton
• ERD → Data Schemas
• Sequence Diagram → Behavioral Flows
• State Machine → Stateful Logic

STRICT RULES:

1. Diagrams are PRIMARY constraints.
2. Preserve dependency direction.
3. Preserve module boundaries.
4. Avoid framework injection unless specified.
5. Report any constraint conflicts.

OUTPUT:

• Architecture-consistent code skeleton
• Mapping Report (Diagram ↔ Code)
• Constraint Violation Report (if any)
"""
```

Fig. 8. Process B: Example of the prompt for code generation from TD.

To meet this requirement, we use DSPy (Declarative Self-Improving Model Programming Language) [27-28] as a structured platform for query optimization. DSPy allows declaratively defining transformation goals, constraints, and evaluation signals, while supporting automated query strategy improvement. Instead of manually iterating query variants, DSPy formalizes query behavior as an optimized program, allowing controlled adaptation based on measurable quality metrics.

In the proposed architecture-mediated pipeline, DSPy performs three main roles:

- stabilization of hint semantics in transformation stages;
- automated optimization of constraint interpretation;
- reduction of hint-induced variability.

With DSPy integration, question tuning moves from an ad hoc empirical process to a systematic optimization procedure aligned with architectural invariants. This approach ensures that diagrams function as basic structural constraints, while DSPy regulates question-level behavior, improving reproducibility, semantic accuracy, and generative consistency.

*G. Verification methods*

*Expert assessment*. Architects and software engineers evaluated the quality and integrity of the unified architecture according to predefined evaluation protocol.

*Cross-platform testing*. The system prototype was tested in various domains, such as database migration, microservices, ETL pipelines and business rule coding, which enabled us to evaluate the generality of the proposed model.

*Comparison with similar methods*. The results were compared with traditional document-based methods with UML and architecture fragments created by architects and generated by modern AI-based tools, such as GitHub Copilot.

*H. Dataset Creation*

To construct the validation dataset, the expert group employed stringent inclusion criteria to ensure structural integrity and traceability across layers of abstraction. Only software systems (Table 2) meeting the following criteria were selected:

- software must be open-source and publicly accessible;
- a complete source code repository must be provided;
- explicit TD (e.g., architecture description, API documentation, developer guidelines) must be provided;
- BD layer or functional descriptions, including use cases, product descriptions, or functional overviews, must be provided.

This selection strategy ensured that each selected system contained sufficient artifacts to support a complete bidirectional transformation cycle: "Code → TD → BD → TD → Code".

By limiting the dataset to systems with transparent source code and multi-layered documentation, we ensured objective traceability, reproducibility of the transformation, and consistent assessment of semantic preservation across layers of abstraction.

TABLE II.  A LIST OF SELECTED SOFTWARE FOR VALIDATION

| Software | Target Criteria | Links (Code, TD, BD) |
|---|---|---|
| LangChain | LLM applications, agents, RAG | https://github.com/langchain-ai/langchain \| https://docs.langchain.com \| https://www.langchain.com |
| Unity ML-Agents | ML agents, reinforcement learning | https://github.com/Unity-Technologies/ml-agents \| ttps://unity-technologies.github.io/ml-agents/ \| https://unity.com/products/machine-learning-agents |
| GenAI_Agents | LLM-based agents, MAS | https://github.com/NirDiamant/GenAI_Agents \| https://github.com/NirDiamant/GenAI_Agents#readme \| https://medium.com/@nirdiamant |
| 500-AI-Agents-Projects | AI agent use cases catalog | https://github.com/ashishpatel26/500-AI-Agents-Projects \| https://github.com/ashishpatel26/500-AI-Agents-Projects#readme \| https://www.analyticsvidhya.com |
| Nekro-Agent | Extensible LLM agent framework | https://github.com/KroMiose/nekro-agent \| https://doc.nekro.ai \| https://nekro.ai |
| Flyway | Database schema migration | https://github.com/flyway/flyway \| https://documentation.red-gate.com/fd \| https://www.red-gate.com/products/flyway |
| pgloader | DB → PostgreSQL migration | https://github.com/dimitri/pgloader \| https://pgloader.readthedocs.io \| https://wiki.postgresql.org/wiki/Using_pgloader |
| Debezium | Change Data Capture (CDC) | https://github.com/debezium/debezium \| https://debezium.io/documentation/ \| https://debezium.io |
| Airbyte | ETL / Data integration | https://github.com/airbytehq/airbyte \| https://docs.airbyte.com \| https://airbyte.com |
| Apache Sqoop | Legacy data migration | https://github.com/apache/sqoop \| https://sqoop.apache.org/docs/ \| https://sqoop.apache.org |
| OpenRewrite | Automated code, config migration | https://github.com/openrewrite/rewrite \| https://docs.openrewrite.org \| https://www.openrewrite.org |
| Ora2PG | Oracle → PostgreSQL migration | https://github.com/darold/ora2pg \| https://ora2pg.darold.net/documentation.html \| https://ora2pg.darold.net |

The experimental set includes applications focused on working with LLMs, which implement the functions of generating code, text, documentation, and dialogue interfaces. These systems are characterized by a multi-layered architecture that combines application logic with models, context storage and prompt management mechanisms, making them typical examples for studying architectural representation.

Another group consists of ML-oriented systems, specifically pipelines for training, evaluating, and running ML models. These systems encompass data processing, model training, validation, and the integration of results with application services, enabling us to verify that the architecture diagram reflects both static and dynamic aspects. Furthermore, it is assumed that some processes were developed using cloud-based tools.

Agent and agent-based systems are considered, in which program logic is implemented through the interaction of autonomous components or agents. Such systems are particularly sensitive to the accurate description of interactions, messaging protocols and behavioral scenarios, making them important test cases for integration diagrams and sequence diagrams.

In addition, these experiments also include systems focused on database operations and data migration, particularly scenarios involving the transfer of templates and code between different database management systems. Such applications enable us to evaluate the extent to which the proposed architecture supports formal descriptions of data structure, dependencies, and transformations, as well as their use in automated code generation and verification.

To create a set of software products used in this study for validation, we used existing open-source software search and selection tools. This search was primarily done on open-source software repositories, mainly on the GitHub platform, which is the well-known largest open-source software repository and is widely used in both industry and academic research. Based on hashtags, keywords and categories, covering areas such as LLMs, ML, AI agents, database migration, data integration, template evolution and ETL pipelines.

Of course, the variety of existing applications is much greater but this approach to validation allows us to assert that the research results are not limited to a specific domain but are of a generalized nature, suitable for application in a wide range of modern software systems.

IV. RESULTS AND DISCUSSION

A. Architecture Frameworks Overview

It is important to distinguish between the unified architectural metamodel and its diagrammatic representations. The metamodel defines the abstract entity types, relationships, constraints, and transformation rules. Architectural diagrams represent concrete visualization instances of this metamodel rather than the metamodel itself.

As part of the framework selection process, a panel of experts composed of experienced software architects and solution engineers with extensive expertise in enterprise and large-scale customer projects examined the architectural practices used in real-world applications. The selection was based not only on academic recognition but also on practical experience in various industry projects, including enterprise systems, cloud-native platforms, AI-based applications, and migration projects.

This expert screening identified over fifty architecture frameworks, modeling approaches, and documentation standards that are actively used in customer projects. These frameworks cover various layers of abstraction, including enterprise architecture, software architecture structuring, DDD, modeling notations, cloud-native architectures, AI/ML system architectures and standards for architecture documentation.

The resulting set reflects accumulated industry experience rather than just theoretical classification, thus ensuring that the proposed unified architectural metamodel is based on practical applicability and validated architectural patterns observed in heterogeneous production environments. Conditionally, we can split them into 7 groups based on the coverability and documentability of a distributed cloud-based enterprise software system Table 3.

TABLE III. ARCHITECTURAL FRAMEWORK GROUPS

| Group of frameworks | Frameworks |
|---|---|
| Enterprise architecture | TOGAF (The Open Group Architecture Framework)<br>Zachman Framework<br>FEAF (Federal Enterprise Architecture Framework)<br>DoDAF (Department of Defense Architecture Framework)<br>MODAF (UK Ministry of Defence Architecture Framework)<br>AGATE / EAGLE / EU Architecture Frameworks<br>EABOK (Enterprise Architecture Body of Knowledge)<br>Gartner Enterprise Architecture Framework<br>PEAF (Pragmatic Enterprise Architecture Framework)<br>MEGAF (Model-Driven Enterprise Architecture Framework) |
| Software architecture | C4 Model (Context, Container, Component, Code)<br>Arc42 Software Architecture Template;<br>Kruchten's 4+1 View Model<br>RUP (Rational Unified Process) Architectural View Set<br>SEI Views and Beyond Method<br>ATAM (Architecture Tradeoff Analysis Method)<br>SAAM (Software Architecture Analysis Method)<br>CBAM (Cost-Benefit Analysis Method)<br>SOMF (Service-Oriented Modeling Framework)<br>OSAF (Open Security Architecture Framework)<br>ISO/IEC/IEEE 42010 Architecture Description Framework<br>V-Model / V-Model XT Architecture Approach |
| Domain-driven and design-centric | DDD (Domain-Driven Design)<br>Hexagonal Architecture (Ports and Adapters)<br>Clean Architecture (Onion Architecture)<br>Microkernel Architecture<br>Event-Driven Architecture (EDA) Reference Frameworks |
| Modelling and representation | UML (Unified Modeling Language)<br>SysML (Systems Modeling Language)<br>BPMN (Business Process Model and Notation)<br>ArchiMate (Enterprise & Application Modeling Language)<br>IDEF (IDEF0, IDEF1x, IDEF3)<br>DMN (Decision Model and Notation) |
| Distributed and cloud-based architecture | CNCF Cloud Native Architecture Framework<br>AWS Well-Architected Framework<br>Azure Architecture Framework<br>Google Cloud Architecture Framework<br>IBM Cloud Architecture Center<br>Netflix OSS Architecture Principles<br>Microservices Architecture Reference Model<br>Twelve-Factor App Architecture |
| AI/ML/LLM-based application architecture | MLflow System Architecture (MLOps)<br>TensorFlow Extended (TFX) Architecture<br>NVIDIA AI Enterprise Architecture<br>OpenAI Reference Architectures for LLM Applications<br>LangChain Modular Architecture<br>DSPy Architectures for Declarative AI Pipelines<br>RAG (Retrieval-Augmented Generation) Architectural Patterns<br>Agent-Based AI System Architecture Frameworks |
| Architecture documentation | ISO/IEC/IEEE 42020 (Architecture Processes)<br>IEEE 1016 (Software Design Description Framework)<br>Software Architecture Document (SAD) templates<br>ADR (Architectural Decision Records) framework |

*Enterprise architecture frameworks with a focus on high-layer business with IT industry alignment.* This group of frameworks defines the overall structure and management of an enterprise, integrating business strategy, organizational processes, information flow, applications and technical infrastructure into a single model. They support long-term planning, capability management, interoperability, and strategic decision-making.

*Software architecture frameworks with a focus on system-layer architectural structuring.* These frameworks focus on the architectural design and analysis of single software systems, describing their components, interactions, quality attributes, limitations, and evolution mechanisms. They provide perspectives, architectural styles, design principles, documentation templates, and evaluation methodologies to support system-layer decision-making. Such frameworks can help engineers track system architecture, domain decomposition, deployment topology, and runtime behavior, while ensuring maintainability, scalability, and modifiability.

*Domain-driven and design-centric frameworks.* The group emphasizes modeling the domain logic, its boundaries and the conceptual integrity of complex systems. Frameworks such as DDD and Clean Architecture advocate for task separation, domain knowledge encapsulation and the establishment of clear boundaries as bounded contexts. These frameworks ensure high component integrity and loosely coupled subsystems by directly coupling architecture with business semantics, providing long-term support. They are particularly important for large, evolving systems with complex business rules.

*Modelling and representation frameworks based on diagrammatic and formal modelling.* These frameworks define visualization and formalization languages, such as UML, BPMN, and ArchiMate, which use standardized notation to describe software and business architecture. They act as an intermediate representation layer between requirements, design and implementation. Architectural diagrams facilitate concept refinement, stakeholder communication, knowledge acquisition, documentation and automated analysis. They are crucial for creating consistent architectural diagrams and providing sufficient justification for system structure and behavior.

*Distributed and cloud-based architectural frameworks.* These frameworks guide the design of scalable, resilient, and maintainable systems deployed in cloud environments or distributed infrastructures. They cover best practices in microservices, container orchestration, reliability engineering, DevOps and operational observability. Exemplifications such as AWS Well-Architected or CNCF provide structured design principles, such as security, scalability, cost optimization, and resilience, that guide the development of cloud systems and ensure their superior operation.

*AI/ML/LLM-based application architecture frameworks.* New, rapidly evolving AI pipelines, model-driven systems, and LLM-based application frameworks address issues related to data flow, training, inference, model management, monitoring and generation, while enhancing search capabilities. These frameworks define reusable architectural patterns for building autonomous agents, dialogue systems, and hybrid AI and software pipelines. They support reproducibility, multi-tiered operations and maintenance (MLOps), scalability, attribution analysis, and appropriate AI practices, all of which are crucial for modern intelligent systems.

*Architecture documentation frameworks.* These frameworks provide a structure and methodology for building architecture documents that are easy to understand, maintain and version control. They establish rules for describing architectural decisions, constraints, quality attributes, component responsibilities and evolution paths. Standards such as IEEE 1016 and ADR ensure that design principles are followed and that architectural knowledge is available throughout the SDLC. They play a crucial role in system tracking, management, auditing and implementation.

## B. Architectural Diagrams

Analyzing the architectural diagrams across investigated frameworks, SMEs provided the results of framework surface examination (Table 4). The columns represent evaluation dimensions capturing documentation capability, traceability, automation readiness, representational expressiveness and empirical impact on code generation quality. Quantitative criteria were normalized using an ordinal scoring scheme (0–5 scale). Descriptive representational characteristics were preserved as qualitative attributes. The Aggregated Capability Index (ACI) was computed as the arithmetic mean of the ordinal metrics.

The comparison highlights that no single architectural framework fully supports the entire SDLC from source code extraction through TD and BD to automated code generation using LLM. Instead, each framework demonstrates strengths at certain layers of abstraction.

TABLE IV.    AGGREGATED FRAMEWORK CAPABILITY INDEX

| Framework | Code Extraction | TD | BD | BD → TD Mapping | TD → Code Traceability | Automation / LLM Info | Code Coverage by Schemas | Impact on Code Generation | Aspects Covered by Diagrams | Diagram Representation Formats | Aggregated Capability Index (ACI) |
|---|---|---|---|---|---|---|---|---|---|---|---|
| C4 Model | 4.2 | 4.8 | 3.4 | 4.3 | 4.1 | 4.9 | 3.2 | 4.8 | Context, Containers, Components, Code, Dynamics | PlantUML, JSON, Markdown | 4.21 |
| Arc42 | 3.4 | 4.9 | 2.3 | 3.5 | 3.6 | 4.2 | 4.9 | 4.2 | Runtime, Building Blocks, Deployment, Cross-cutting Concerns | Markdown, AsciiDoc, PlantUML | 3.88 |
| DDD | 2.4 | 4.1 | 4.3 | 4.4 | 3.7 | 3.6 | 2.4 | 3.8 | Domain Model, Context Map, Bounded Contexts | UML, JSON, Text | 3.59 |
| CNCF / Microservices | 2.6 | 4.2 | 2.4 | 2.5 | 4.3 | 3.8 | 4.8 | 3.9 | Runtime Topology, Services, Deployments | YAML (K8s), JSON, Mermaid | 3.56 |
| ArchiMate | 4.3 | 4.0 | 1.4 | 2.3 | 4.5 | 1.6 | 4.9 | 2.8 | Business, Application, Technology, Motivation, Implementation | ArchiMate XML, GraphML, Visual Tools | 3.23 |
| UML | 2.3 | 4.0 | 4.1 | 4.2 | 2.3 | 2.4 | 2.3 | 2.9 | Class, Sequence, Deployment, Activity | PlantUML, XMI, SVG | 3.06 |
| TOGAF | 1.4 | 4.2 | 4.9 | 4.8 | 1.5 | 2.1 | 1.3 | 1.9 | Business, Information, Application, Technology, Motivation | ArchiMate XML, BPMN, Proprietary Tools | 2.76 |
| Kruchten 4+1 | 2.1 | 3.2 | 2.4 | 2.3 | 2.2 | 3.1 | 2.0 | 2.6 | Logical, Process, Development, Physical, Scenarios | UML, Text | 2.49 |
| SAAM / ATAM | 1.3 | 3.1 | 3.3 | 4.1 | 2.4 | 2.2 | 0.6 | 1.8 | Quality Attributes, Scenarios | Text, Spreadsheets, Templates | 2.35 |

The frameworks were ranked according to the Overall Capability Index (ACI), and the results showed that structure-oriented, machine-readable architectural approaches held a significant advantage.

The C4 model demonstrates the best overall balance in today's code-centric workflows based on LLM. It breaks down the structure into context, containers, components, and code representation, which aligns well with the interpretation of a system for LLM. C4 supports powerful code retrieval capabilities, excellent documentation writing abilities, and clear traceability between architectural views and implementation artefacts. Its concise and clear diagrams and structure make it easy to serialize into machine-readable formats such as PlantUML or JSON, which explains its significant improvement in code generation quality.

TOGAF excels in BD and enterprise-layer mapping, but contributes little to direct code extraction or automation. Its primary value lies in governance, capability planning, and policy coordination, rather than implementation details. Therefore, TOGAF is best suited as an upstream business framework, providing context and constraints, but must be used in conjunction with more code-oriented frameworks to support technology generation tasks.

ArchiMate sits between the business layer and the technology layer, providing a structured perspective for these layers and facilitating the harmonization and unification of enterprise architecture. However, the layer of abstraction limits its use in generating detailed code or performing template-layer coverage. ArchiMate can improve understanding of the structure, but its impact on the quality of automated code is limited.

Arc42's uniqueness lies in its focus on TD frameworks rather than modeling languages. Its advantage lies in generating comprehensive, well-structured technical descriptions covering execution-time behavior, deployment, and cross-cutting concerns. These descriptions are invaluable input for LLM, significantly improving document quality and subsequent code generation efficiency. However, Arc42's direct support for business modeling is relatively limited.

The Kruchten 4+1 representation model is very effective for scenario-based architectural thinking and stakeholder communication, but its representations focus on description rather than work. This limits automation and traceability, making it less suitable for LLM-driven code generation processes.

DDD achieves strong conceptual consistency between business logic and code through bounded contexts and domain models. While it doesn't directly support code abstraction, it improves semantic clarity and, when used in conjunction with structured frameworks, positively impacts code generation.

UML remains powerful in detailed structural and behavioral modeling and offers excellent schema-layer coverage. However, its weak business semantics and verbose descriptions reduce its effectiveness in achieving comprehensive automation in the absence of additional context.

Ultimately, CNCF and the microservices model are very effective in terms of runtime and deployment representation, especially in cloud systems. They greatly facilitate code consistency at the infrastructure layer, but perform poorly in terms of BD.

Overall, the comparison results show that the combination of C4 and Arc42, with selective enhancements using DDD or ArchiMate, provides the most effective framework for software development using LLM, balancing business understanding, technical clarity, and automation potential. Based on the provided expertise of SMEs, we have constructed the most optimal set of architectural diagrams mixing from different frameworks (Table 5).

TABLE V. A Set of Chosen Diagrams for Investigation

| Diagram Name | Primary Framework / Origin | Formal Role in Unified Metamodel |
|---|---|---|
| Business Context Diagram | C4 (C1), TOGAF, ArchiMate | Defines system boundaries, external actors, upstream/downstream systems, and high-level semantic scope. Serves as a hallucination-reduction constraint and semantic anchoring mechanism for LLM reasoning. |
| Business Capability Map | TOGAF, ArchiMate | Represents stable business abilities independent of implementation. Enables capability-driven decomposition, prioritization, and invariance of business intent across regeneration cycles. |
| Domain Model Diagram | DDD, UML | Formalizes domain semantics via entities, value objects, aggregates, and relationships. Acts as the primary semantic fidelity stabilizer controlling naming, data meaning, and API payload consistency. |
| Business Process Diagram | BPMN | Specifies causal and temporal business logic, workflows, roles, and decision points. Functions as a behavioral constraint layer guiding orchestration, sequencing, and test-case derivation. |
| DDD Context Map | DDD | Establishes bounded contexts, semantic boundaries, and integration relationships. Provides strong guardrails for microservice partitioning and cross-context dependency control. |
| CQRS Diagram | CQRS / Architectural Pattern | Formalizes separation of read/write responsibilities and storage models. Functions as a pattern-level constraint preventing logical mixing in generated systems. |
| Event-Driven Diagram | Event-Driven Architecture (EDA) | Specifies asynchronous flows, messaging channels, handlers, retries, and idempotency constraints. Serves as a concurrency and decoupling stabilizer. |
| Clean / Onion Architecture Diagram | Clean Architecture | Encodes dependency direction rules and abstraction boundaries. Functions as a structural invariance regulator preventing architectural drift during regeneration. |
| System Container Diagram | C4 (C2) | Defines major deployable units, services, repositories, and APIs. Serves as the primary structural decomposition mechanism linking business capabilities to executable system units. |
| Component Diagram | C4 (C3), Arc42, UML Component | Specifies internal modular decomposition, responsibilities, interfaces, and dependency direction. Acts as the principal constraint model for module generation and dependency invariance. |
| Deployment / Infrastructure Diagram | C4 Deployment, Arc42 Runtime, CNCF | Maps logical architecture onto execution environments. Enables infrastructure-as-code (IaC) synthesis and runtime feasibility validation. |
| Integration / API Interaction Diagram | C4 Dynamic, UML Sequence | Captures communication semantics, contracts, protocols, events, and dependency flows. Reduces integration ambiguity and stabilizes interface synthesis. |
| Strangler Migration Diagram | Evolutionary Architecture Pattern | Models incremental modernization and transformation boundaries. Functions as an evolutionary constraint controlling safe regeneration and migration stability. |
| Class / Module Structure Diagram | UML Class Diagram | Defines static code structure, abstractions, polymorphism, and module relationships. Serves as the primary target for deterministic code skeleton generation. |
| Sequence / Interaction Diagram | UML Sequence Diagram | Represents runtime execution logic and causal interaction flows. Reduces behavioral ambiguity and stabilizes orchestration logic generation. |
| Data Model / Schema Diagram | ERD, UML | Specifies persistent structures, schemas, keys, and relationships. Acts as the invariance layer for storage consistency and migration stability. |
| Runtime Topology / Operational Diagram | CNCF / Cloud-Native Architecture | Represents live operational behavior including service instances, queues, caches, scaling groups, resilience mechanisms, and SLO-relevant links. |
| State Machine Diagram | UML State Machine | Formalizes lifecycle states, transitions, guards, and actions. Supports reliable synthesis of stateful logic and edge-case handling. |

Each diagram type is selected based on its ability to reduce ambiguity, preserve semantics across abstraction layers, and support bidirectional transformations between BD, technical architecture, and source code.

*Business and Domain Understanding.* This layer explains the significance of the system's existence and the business problems it solves, and is unrelated to the choice of technology. The key diagrams of the high-layer are formed as follows:

- **Business Context Diagram** - this diagram defines the system boundaries, stakeholders, users, and external systems. Within the framework of SDLC, it serves to define the scope and prevent misconceptions by clearly defining the internal and external aspects of the system.
- A **Business Capability Map** represents an organization's performance in a stable form of business capabilities. It separates business intent from execution processes and provides a consistent semantic foundation for lifecycle management, allowing it to withstand architectural and technological changes.
- **Domain Model Diagram** uses a domain language to define core business concepts, entities, value objects, and relationships, which is essential for LLM-based architectures because it binds code and APIs to common business semantics rather than syntactic structures.
- **Business Process Diagram** describes workflows, roles and decision points. It transforms static functionality into actionable business behaviors and provides causal and progressive logic for SDLC, enabling seamless integration into orchestration, service and agent workflows.
- **DDD Context Map** defines the process of decomposing system semantics into distinct contexts and clearly establishes domain boundaries, context responsibilities, and their integration relationships. In the SDLC, they serve as a mechanism for maintaining conceptual integrity, preventing unintended coupling, and stabilizing the mapping between business semantics and system architecture.

*System Architecture.* This layer explains the system's structure and usage to achieve business objectives. The key architectural diagrams are:

- **The CQRS diagram** defines the structural separation between responsibility for instructions and requests, including the associated data flow and memory model. Within the SDLC, it helps prevent accidental confusion between change logic and selection logic, provides clear boundaries of responsibility, improves scalability decisions, and stabilizes LLM-based code generation through explicit constraints on the coding architecture.
- **Event-driven diagram** defines asynchronous interactions, message channels, event sources, event consumers, handlers, and delivery semantics (retry and idempotence rules). In the SDLC, it is used to formalize distributed communication patterns, clarify time dependencies between components, and reduce ambiguity in the behavior of distributed systems, thereby improving the consistency, reliability, and correctness of the orchestration logic generated by the LLM.
- **Clean/Onion Architecture Diagram** defines the rules for dependency direction, abstraction boundaries, and hierarchical organization of layers in a system. Within the SDLC, it serves to enforce architectural invariants, prevent miscoupling between implementation details and domain logic, and stabilize regeneration processes by explicitly encoding structural constraints that guide both human thinking and LLM-based transformations.
- **System Container Diagram** illustrates the fundamental deployment units, applications, services, repositories and APIs and their interactions. For lifecycle management, this is the primary decomposition method for defining codebases, repositories, and service boundaries.
- **Component Diagram** divides a container into internal components or modules and clearly defines their responsibilities and dependencies. It details issues such as connectivity, coupling, and responsibility allocation.
- **Deployment/Infrastructure Diagram** displays software elements on physical or cloud infrastructure (hosts, containers, Kubernetes nodes, networks). It is required for configuring specific environments and analyzing scalability, reliability, and runtime constraints.
- **Integration/API Interaction Diagram** describes the communication patterns between components or external systems, including APIs, messages, and events. It is particularly important for SDLC when creating interface contracts, integration tests, and orchestration logic.
- The **Strangler Migration Diagram** simulates the gradual modernization process of legacy systems by defining the transition and coexistence boundaries between legacy components and new elements. In the SDLC, it serves as an evolutionary constraint mechanism, guiding controlled replacement, interface stability, and dependency redirection. By formalizing migration paths, the diagram reduces modernization risks, prevents architectural drift, and supports stable regeneration strategies within evolving systems.

*Developer-Layer Architecture.* This layer captures how the system is implemented and is closest to the source code. The key architectural diagrams at the:

- **Class/Module Structure Diagram** used to identify categories or modules, their attributes, methods, and static relationships. This diagram can be directly converted into a code framework and is crucial for understanding patterns.
- **Sequence/Interaction Diagram** depicts the step-by-step interactions during the execution of a specific use case. It provides a logical execution path and time sequence for the SDLC, thereby improving the correctness of the resulting processes and business logic.
- **Data Model/Schema Diagram** represents the database schema, tables, keys, and relationships. This diagram ensures the correct generation and migration of the retention layer.
- **Runtime Topology/Operation Diagram** describes the behavior in a production environment, including expansion groups, queues, caches,

service networks, and operational connections. It supports the creation of deployment descriptors, configuration of monitoring mechanisms, and resilience.
- **State diagrams** formalize the dynamic behavior of a system by defining SFS, transitions, triggering events, protection conditions, and concluding actions. Unlike structure diagrams, they capture the temporal evolution of a system. Within the SDLC, this diagram serves to model behavioral invariance, thus ensuring the consistency, predictability, and verifiability of the execution logic. By explicitly representing state dependencies and transitions, state diagrams reduce ambiguities in behavioral interpretation and support the reliable synthesis of stateful logic, exception handling, and constraints.

How to Choose the Right Diagram? Diagram selection should be targeted and not an attempt to cover everything. Some suggestions for choosing the right set of architectural diagrams:

- To clarify scope and prevent hallucination: start with Business Context.
- To align code with business meaning: use Domain Model and Capability Map.
- To structure services and repositories: rely on System Container and Component diagrams.
- To generate APIs or workflows: add Integration and Sequence diagrams.
- To generate schemas or migrations: include Data Model diagrams.
- To generate deployment and configs: use Deployment and Runtime Topology diagrams.

Some diagrams are considered by different architectural frameworks (Table 5), thus it is necessary to distinguish between the dominant framework and the auxiliary ones. This is necessary to avoid conceptual confusion, duplication of artifacts and contradictions in the interpretation of architectural elements.

The dominant framework defines the main coordinate system - terminology, layers of abstraction, representation structure and rules for building diagrams. Auxiliary frameworks can be used to detail individual aspects, for example, behavioral or infrastructure, but they must be consistent with the basic model and not violate its logic.

A clear separation of framework roles ensures the integrity of the architectural description, facilitates tracing between preferred representations, and maintains consistency between the conceptual, logical, and physical layers of the architecture.

### C. Refining the List of Architectural Diagrams based on Requirements to Architecture Design and Architectural Patterns

Let's consider a use case when you should refine a set of architectural diagrams of unified architectural metamodel.

The choice of architectural metamodel directly influences which diagrams are most suitable and effective in describing and understanding the system architecture. Each architectural model defines a specific set of structural elements, interaction types, and dependencies between components, thus requiring appropriate visualization methods. For example, component diagrams or container diagrams are more suitable for multi-tiered architectures, as they can reflect the hierarchical structure and direction of dependencies; while for event-driven architectures or microservice architectures, event flow diagrams, sequence diagrams, or deployment diagrams are crucial.

Therefore, each architectural pattern has a corresponding diagram type that optimally reflects its essence, key constraints, and interaction mechanisms. Appropriate graphical representations deepen the understanding of architectural solutions, promote effective communication between stakeholders, and ensure consistency between the conceptual model and its implementation.

Meanwhile, creating such diagrams also requires meeting some general requirements: they must reflect the correct dependencies between components, conform to the chosen layer of abstraction, support different perspectives (structure, behavior, implementation-related), and remain consistent with the actual system state. However, despite existing methods and recommendations, some unresolved issues remain in both academic and practical contexts. These issues include:

- diagrams are difficult to adapt to actual architectures and code libraries;
- a lack of universal standards applicable to different patterns;
- a lack of empirical evaluation of the effectiveness of certain visualization types for specific architectural contexts.

Also, the following research problems are still unresolved:

- Technologies for automatically generating correct diagrams for specific patterns are not yet mature. Existing methods may lag behind the actual code/system architecture and cannot always account for all patterns.
- Aligning diagrams at different layers of abstraction (from C4 to UML/code patterns) remains difficult, especially for large systems.
- The human factor in visualization. While diagrams are effective, their application depends significantly on the practical experience of developers and architects, which complicates standardization.

Based on the analysis of design principles, the following are the most important requirements for diagram processing:

- Accurately represent dependencies between modules/components, for example: Single Responsibility Principle (SRP) and Dependency Injection Principle (DIP).
- Highlight the roles of interfaces and implementations, for example: Interface Structure and Implementation Structure (ISP) and the Hexagonal pattern.
- Demonstrate the allocation of responsibilities, for example: applicable to patterns that improve system integration.
- Communicate the behavior of the system/dependencies over time, for example: Event-Driven Design Principles and Component Requirement Response Structure (CQRS).
- Show the hierarchy of abstractions such as is in C4: "Context → Container → Component → Code".

Some requirements to diagrams according to architecture design principles with examples are shown in Table 6.

TABLE VI. RELATIONSHIPS BETWEEN DESIGN PRINCIPLES AND ARCHITECTURAL DIAGRAMS VIA REQUIREMENTS

| Design Principle | Requirements | Similar Diagram |
|---|---|---|
| Layered / N-Tier | Hierarchy of responsibilities and their dependencies | C4 Container/Component, UML Component |
| Hexagonal / Ports and Adapters | Connection ports/adapters – kernel interaction | UML Component, Layered/Ports diagram |
| CQRS / Event-Driven | Commands and requests, events, and asynchronous threads | Sequence Diagram, Event flow diagram |
| Microservices | Services and their APIs/integration | Deployment Diagram, C4 Context |
| MVC / MVVM | Control distribution and representation | UML Class/Sequence |
| Dependency Injection / DIP-oriented | Implementation of external interfaces | Component Dependency Diagram |
| Facade / KISS-oriented | Simplified subsystem interfaces | UML Component, Facade Package Diagram |

Recent systemic research findings and empirical findings of architectural patterns consider the following requirements for architectural diagram:

- Components and their relationships are represented based on the selected pattern, for example, dependencies, subsystems, event channels, services, modules, etc.
- Various views are supported—structural, behavioral and deployment views with multiple perspectives.
- Users should understand not only the architecture but also the underlying solution (design principles).
- The abstraction layers of stakeholders, such as architects, developers, analysts, and DevOps engineers, are taken into account.

These requirements are crucial for successful communication, decision-making and architectural support. Consistency with the architecture documentation is ensured to avoid discrepancies between diagrams and code.

Architectural principles express the fundamental forces and constraints that shape a system, such as maintainability, scalability, loose coupling, and the ability to evolve. They clarify why certain properties are necessary. Design and architectural patterns put these forces into practice by providing concrete structural or behavioral solutions, thus realizing these principles. Principles therefore define intention and direction, while patterns translate these intentions into realized architectural structures.

Architectural patterns become observable and verifiable through diagrams. Diagrams formally visualize components, dependencies, processes, and boundaries, allowing one to assess whether the chosen pattern truly adheres to its fundamental principles. In this sense, diagrams can confirm structural compliance and communicate architectural decisions to stakeholders (Table 7), thereby improving transparency, traceability and consensus within the development team.

TABLE VII. DESIGN PRINCIPLES, ARCHITECTURAL PATTERNS AND ARCHITECTURAL DIAGRAMS

| Design Principle | Architectural Patterns | Target Architectural Diagram | Major Assignment |
|---|---|---|---|
| SRP (Single Responsibility) | Facade, Adapter, MVC, Layered Architecture, Modular Monolith | C4 Component Diagram, UML Component Diagram, Layered Architecture Diagram | One component = one reason to change |
| OCP (Open/Closed) | Strategy, Factory Method, Template Method, Plugin Architecture, Microkernel | UML Class Diagram (Strategy), Plugin Architecture Diagram, Sequence Diagram (runtime extension) | Extend behavior without modifying core |
| LSP (Liskov Substitution) | Strategy, State, Decorator, Polymorphism-based Hierarchies | UML Class Diagram with inheritance, State Machine Diagram, Behavioral Test Matrix (architecture quality artifact) | Subtypes preserve behavioral contracts |
| ISP (Interface Segregation) | Ports and Adapters (Hexagonal), Role Interfaces, Proxy | Hexagonal Architecture Diagram, UML Interface Diagram, C4 Component Diagram (port-based view) | Clients depend only on minimal interfaces |
| DIP (Dependency Inversion) | Dependency Injection, Clean Architecture, Onion Architecture, Hexagonal Architecture | Clean Architecture Circles Diagram, Dependency Direction Diagram, Composition Root (wiring) Diagram | Dependencies point to abstractions, not implementations |
| DRY ("Don't Repeat Yourself") | Template Method, Builder, Shared Kernel (DDD), Code Generation | Template Method UML Diagram, Feature-to-Module Mapping Diagram, Code Generation Pipeline Diagram | Centralized logic, avoid duplication |
| Separation of Concerns (SoC) | Layered Architecture, MVC, CQRS, Event-Driven Architecture | CQRS Diagram, Event-Driven Architecture Diagram, Layered Architecture Diagram | Distinct concerns isolated structurally |
| Low Coupling / High Cohesion | Mediator, Observer (Pub/Sub), Domain Events, Microservices, Bounded Context (DDD) | DDD Context Map, Event Storming Diagram, Service Interaction Diagram | Reduce direct dependencies, group related behavior |
| KISS ("Keep It Short and Simple") | Facade, Opinionated Architecture, Convention over Configuration | Facade UML Diagram, Simplified API Surface Diagram, "Happy Path" Sequence Diagram | Reduce accidental complexity |
| YAGNI ("You Aren't Gonna Need It") | Strangler Pattern, Modular Monolith, Evolutionary Architecture | Strangler Fig Migration Diagram, Roadmap Architecture View, Evolutionary Fitness Function Diagram | Implement only required features, evolve safely |

*D. Formalization of the Unified Architectural Metamodel*

To enable system transformation between business requirements, TD, and source code, this paper presents a unified architectural metamodel that formalizes architectural knowledge into a structured, machine-interpretable representation.

A metamodel defines a set of basic entities, relationships, and constraints that collectively describe the operation of a software system at multiple layers of abstraction. The metamodel acts as an intermediary layer between manually generated architectural descriptions and the SDLC-based generative process.

The metamodel defines the following main entity types:

- *BusinessCapability* – represents a business function or value-delivering capability.
- *BusinessProcess* – describes a sequence of activities realizing a business capability.
- *System* – a deployable or logical software system providing functionality.
- *Container* – an independently deployable unit within a system (e.g., service, application).
- *Component* – an internal structural unit of a container responsible for a specific function.
- *Agent* – an autonomous or semi-autonomous computational entity, often driven by LLMs.
- *DataStore* – a persistent storage entity, including databases, vector stores, or file systems.
- *API / Interface* – a formal interaction point between components or systems.
- *DeploymentNode* – an execution environment such as a VM, container runtime, or cloud service.

The unified architectural metamodel supports a multi-layer hierarchy of abstractions (Table 8) that enables consistent transformation across business, architectural, operational, implementation, and evolutionary perspectives. These layers form a structured continuum from intent to executed systems.

Clearly defining abstraction layers in a single architectural metamodel is not merely a classification task, but a fundamental mechanism to ensure semantic consistency, traceability, and automation throughout the SDLC. Each layer of abstraction represents a unique perspective of the system and takes into account different stakeholders, decision-making domains, and transformation needs.

At the *business layer*, architecture embodies intent, value creation, and corporate goals. This layer defines the purpose of the system and the business problem it solves. Without this semantic anchor, subsequent architectural representations may be technically consistent but strategically divergent. In AI-based systems, especially those with Logical Layer Models (LLMs), preserving business semantics is critical to avoiding context drift and functional inconsistencies.

TABLE VIII. ABSTRACTION LAYERS IN THE UNIFIED ARCHITECTURAL METAMODEL

| N | Abstraction Layer | Typical Diagram Types | Core Purpose | Metamodel Entities Covered | Role in Transformation Cycle |
|---|---|---|---|---|---|
| 1 | Business | Business Context, Capability Map, Business Process | Define organizational intent and value delivery | BusinessCapability, BusinessProcess, Stakeholder, Role | Starting point for Biz → Tech mapping |
| 2 | Business / Conceptual | Domain Model, DDD Context Map | Formalize domain semantics and bounded contexts | DomainEntity, ValueObject, Aggregate, BoundedContext | Preserve semantic consistency Biz ↔ Tech |
| 3 | Business / System | Context Map, Capability–System Mapping | Translate business capabilities into system boundaries | System, CapabilityMapping, IntegrationPattern | Aligns business intent with system decomposition |
| 4 | System | C4 Container, Component Diagram | Logical system decomposition | System, Container, Component, API, Interface, Agent | Core anchor for Tech Doc ↔ Code traceability |
| 5 | System / Pattern | CQRS, Event-Driven, Sequence (high-layer) | Operationalize architectural patterns | Command, Query, Event, Handler, Policy | Enforce structural, behavioral separation |
| 6 | System / Structural | Clean/Onion, Layered Architecture | Enforce dependency direction, architectural invariants | Layer, DependencyRule, AbstractionBoundary | Prevent architectural erosion during regeneration |
| 7 | System / Runtime | Deployment Diagram | Map software to execution environments | DeploymentNode, Runtime, Cluster, Network, ContainerInstance | Connect architecture to infrastructure-as-code |
| 8 | Runtime | Runtime Topology / Operational Diagram | Model live operational behavior | ServiceInstance, Queue, Cache, ScalingGroup, SLO | Ensure production-layer consistency |
| 9 | Implementation | UML Class, ERD / Data Model | Define static code and schema structures | Module, Class, Function, Method, Schema, Table | Primary target for Tech Doc → Code synthesis |
| 10 | Implementation / Behavioral | Sequence Diagram | Model runtime execution flows per use case | Interaction, Call, Message, ExecutionPath | Reduce ambiguity in generated logic |
| 11 | Behavioral | State Machine Diagram | Formalize lifecycle, state transitions | State, Transition, Guard, Action | Validate dynamic correctness |
| 12 | Evolutionary | Strangler Migration Diagram | Control incremental modernization | LegacySystem, NewSystem, MigrationStep, RoutingRule | Manage safe transformation over time |

The *Business/Conceptual layer* and the business/system layer ensure that domain meaning is not lost during the transition to the technical architecture. These layers formalize domain entities, bounded contexts, and system mapping capabilities. Their importance lies in maintaining semantic accuracy during the transformation from abstract requirements to structural design.

The *System layer* provides the primary means of decomposing the architecture into containers, components, services, and agents. This layer is critical for traceability between documentation and implementation. It forms the structural foundation of the system and is often the primary layer for LLM-enabled code generation because it balances abstraction and concreteness.

The *System/Pattern layer* and the *System/Structural layer* introduce various architectural constraints and limitations. The template layer puts design principles into practice, while the structure layer defines dependency directions and module boundaries. These layers are crucial for preventing architectural erosion and ensuring the automatic generation of expected design invariants.

The *System/Runtime layer* and the runtime layer ensure the tight integration of the architecture into the actual operating environment. They take into account deployment topology, scalability, infrastructure dependencies, and runtime interactions. Their inclusion is essential for modern distributed and cloud systems, as structural correctness alone does not guarantee operational feasibility.

At the *Implementation layer*, the architecture can be directly converted into a raw code structure. This layer is the primary target for synthesis and regeneration in the AI pipeline. However, without an upstream abstraction layer, the implementation risks becoming an isolated structure lacking semantic consistency.

The *Implementation/Behavior layer* and the behavior layer formalize dynamic execution logic and state transitions. They are especially important for validating correctness, ensuring workflow consistency, and modeling lifecycle constraints.

Finally, the *Evolution layer* introduces temporal abstraction. The architecture is not static, but constantly evolving.

By modeling migration paths, transitions from legacy systems to other systems, and modernization strategies, the metamodel supports the safe and controlled evolution of the system, which is crucial for long-running enterprise systems.

Together, these abstraction layers form a multidimensional architectural structure. Their explicit separation enables bidirectional traceability, improves the robustness of automation, reduces ambiguities, and ensures consistency, verifiability, and machine interpretation during transformations between business logic, architecture, runtime environment, and code. The mapping relationships between these layers are formally defined, thus achieving traceability from business requirements to generated code and reverse tracing.

The extended abstraction structure transforms the metamodel from a simple three-layer hierarchy into a multi-dimensional architectural lattice, where:

- Business layers define intent.
- Structural layers enforce constraints.
- Pattern layers operationalize forces.
- Runtime layers reflect execution reality.
- Implementation layers encode realization.
- The evolutionary layer governs change.

This layered structure enables consistent, traceable, and machine-interpretable transformation across the entire AI-driven SDLC cycle. Mapping (Table 9) all things together, we are able to define a required set of diagrams per each use case.

TABLE IX. MAPPING ARCHITECTURAL DIAGRAMS ON PRIMARY ABSTRACTION LAYERS

| Architectural Diagrams | Primary Abstraction Layer | LLM / Automation Role |
|---|---|---|
| Business Context Diagram (C4 C1) | Business | Strong constraint to reduce hallucinations; improves relevance and completeness |
| Business Capability Map | Business | Enables capability-driven decomposition and prioritization |
| Domain Model Diagram | Business / Conceptual | Improves naming, data semantics, API payload correctness |
| Business Process Diagram (BPMN) | Business | Guides workflow/agent orchestration; supports test-case derivation |
| DDD Context Map | Business / System | Strong semantic guardrails for microservice boundaries and interfaces |
| System Container Diagram (C4 C2) | System | High value for repository/service scaffolding and dependency control |
| Component Diagram (C4 C3 / UML Component) | System | Drives module generation, interface definitions, dependency injection patterns |
| Integration / API Interaction Diagram | System | Generates contracts, clients, adapters; reduces integration defects |
| CQRS Diagram | System / Pattern | Helps LLM generate correct separation of responsibilities and storage models |
| Event-Driven Diagram | System / Pattern | Guides generation of messaging code, idempotency, retries, handlers |
| Clean / Onion Architecture Diagram | System / Structural | Strong constraint mechanism: reduces architectural drift in generated code |
| Deployment / Infrastructure Diagram | System / Runtime | Enables IaC/K8s manifest generation; improves realism of generated configs |
| Runtime Topology / Operational Diagram | Runtime | Drives observability, resilience patterns, scaling configuration |
| Class / Module Structure Diagram (UML Class) | Implementation | High precision for code skeletons and refactoring guidance |
| Sequence / Interaction Diagram | Implementation / Behavioral | Improves behavioral correctness; supports test and contract generation |
| State Machine Diagram | Behavioral | Supports reliable generation of stateful logic and edge-case handling |
| Data Model / Schema Diagram (ERD) | Implementation | Essential for DB schema generation, migration scripts, ORM mapping |
| Strangler Migration Diagram | Evolutionary | Guides LLM to generate changes incrementally; reduces migration risk |

Although architectural diagrams are characterized by structured formality and multi-layer design, they do not guarantee complete coverage of system knowledge. Every architectural diagram is an abstraction, and every abstraction by its nature involves intentional omission of some information. Therefore, difficulties in conveying semantic context can arise when switching between different diagram types or abstraction layers, for example: "business process diagram → component diagram → code".

*E. Compensation of Content Losses*

Architectural drawings are an important design tool, but they no longer pass entirely in today's context. Over the past thirty years, numerous negotiations have likely arisen. In the early stages of an LLM program, there's no need to consider "business → architecture → code"—or rather, no need to focus on architectural innovation and resident planning. Obviously, architectural diagrams cannot fully convey all the details of a project. Thus, the content losses may be caused by the following reasons:

- Different layers of abstraction (business semantics are implied in the technical structure).
- Different diagram focuses (structure diagram, behavior diagram, or infrastructure diagram).
- Implicit assumptions (architectural principles not formally defined in the diagrams).
- Incomplete coverage of non-functional requirements.
- Lack of clear mapping relationships between different diagram elements.

For example: a domain model can describe concepts in detail but not explain their boundaries; a deployment diagram can depict the environment but not the resource allocation; a sequence diagram can provide the foundation. This means that our diagrams are not automatically semantic.

When using LLMs for automated document generation, architecture refactoring, or code synthesis, the integrity and clarity of the context are crucial. Incomplete, non-standardized, or implicit contexts can lead to semantic distortions, overinterpretations of architectural schemas, faulty or inconsistent relationships between components in the original model, and violations of structural constraints and dependencies.

LLM focuses on developing the latest information that can be quickly and effectively integrated into real-world applications. Each model's architecture diagram cannot cover all relevant use cases, as it is inherently a flawed model. It includes some cutting-edge technologies related to network data transmission and data delivery.

If the architecture diagram does not fully cover all aspects of the system, context-expanded hints are needed, as follows:

- Explicitly specify:
  - Architectural principles (SRP, DIP, etc.);
  - Constraints ("Services should not directly access other service repositories");
  - Non-functional requirements;
  - Patterns (CQRS, EDA, etc.).
- Specify the mapping relationships between layers:
  - Which business opportunity corresponds to which container?
  - Which domain entity corresponds to which table?
- Add descriptions regarding allowed transformations:
  - What can be changed?
  - What should remain unchanged?

In this context, we are not discussing injection in the traditional sense, but rather controlled, context-aware extensions to prompts. The following points should be noted:

- If the LLM uses the architecture diagram as its primary source of information, additional text in the prompts may alter the interpretation of the architecture.
- Uncontrolled insertions can lead to inconsistencies between the architecture diagram and the textual instructions.

Therefore, it is recommended to separate:

- canonical architectural context (architectural diagram/JSON model);
- instructions for the generation operation;
- explicitly specified what remains unchanged.

All elements of the metamodel can be serialized into structured formats (Mermaid, JSON, PlantUML, DSL). These representations preserve semantic information while being compact enough to be included in LLM prompts, making the metamodel suitable for use at both the architecture and context layers of the generated system.

*F. Expert-based Initial Results of Assessment*

Let's resume the conditions for experiments. To evaluate the practical effectiveness of the proposed unified architecture metamodel, we conducted a controlled comparative experiment with two transformation processes: text-centric transformation and architecture extension transformation based on structured diagrams.

Twelve representative open-source systems, selected from LLM-based agents, machine learning pipelines, database migration tools, and ETL platforms, served as the evaluation objects. Each system underwent a complete bidirectional regeneration cycle: "Code → TD → BD → TD → Regenerated code".

All evaluation automatically calculated metrics were first transformed into dimensionless normalized scores within the [0–1] interval. A uniform linear mapping was subsequently applied to obtain ordinal values within the [0–5] range. This normalization strategy preserves proportional metric differences while ensuring interpretability and cross-metric comparability.

As it was already mentioned before, in parallel for each experiment, we recorded seven architecture quality metrics on a predefined scale of 0 to 5 for evaluation by SMEs. The evaluation was conducted independently by three experienced senior software architects. For example, the results of experiments with Ora2PG are shown in Table 10.

TABLE X. AGGREGATED RESULTS FOR ORA2PG ESTIMATION

| Metrics | Process A | | Process B | |
|---|---|---|---|---|
| | Calculated | SMEs | Calculated | SMEs |
| Completeness | 3.2 | 3.0 | 4.6 | 4.5 |
| Semantic Fidelity | 2.6 | 2.8 | 4.1 | 4.2 |
| Consistency | 2.7 | 2.9 | 4.4 | 4.6 |
| Traceability Coverage | 1.9 | 2.1 | 4.3 | 4.4 |
| Machine Readability | 2.1 | 2.3 | 4.9 | 4.8 |
| LLM Constraint Effectiveness | 2.4 | 2.6 | 4.5 | 4.7 |
| Code Pattern Coverage | 3.3 | 3.2 | 4.0 | 4.0 |

During experiments with Process B we've got 5 diagrams produced as explicit diagram artifacts:

- Business Context Diagram (Fig. 9);
- Business Capability Map (Fig. 10);
- System Container Diagram (Fig. 11);
- Integration/API interaction view (Fig. 12)
- Strangler Migration Diagram (Fig. 13)

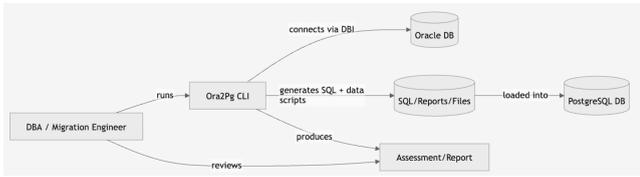

Fig. 9. Process B: Business Context Diagram for Ora2PG.

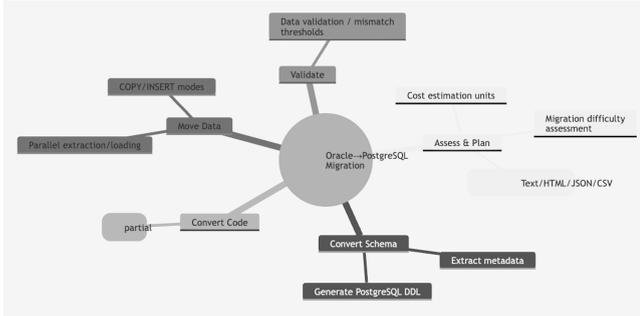

Fig. 10. Process B: Business Capability Map for Ora2PG.

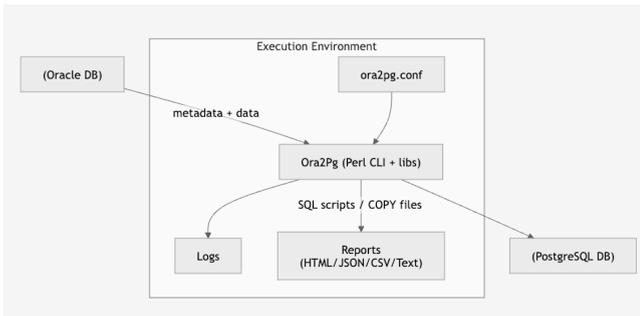

Fig. 11. Process B: System Container Diagram for Ora2PG.

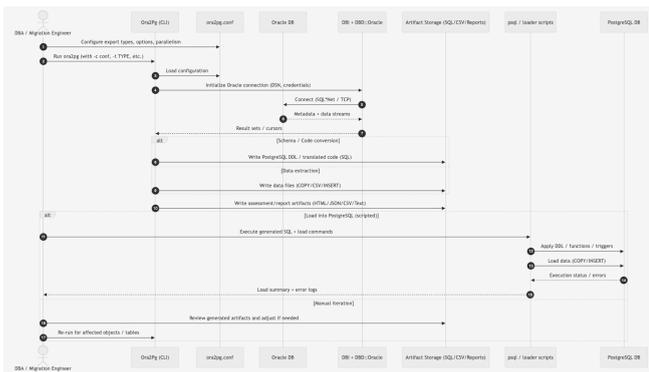

Fig. 12. Process B: Integration/API interaction view for Ora2PG.

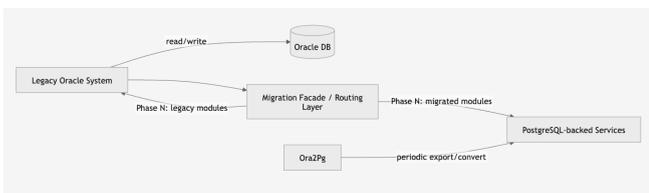

Fig. 13. Process B: Integration/API interaction view for Ora2PG.

And in the strict definition of process B (Table 9), the pipeline could use up to 18 diagram types, but for Ora2Pg only the subset that was clearly justified from the context of the public repository/documentation.

The summary results (Table 11) for heterogeneous software systems show that architecture-driven transformations consistently outperform text-based pipelines. The most significant improvements relate to traceability reconstruction, structural consistency, and machine interpretation. This confirms the role of diagrams as generative constraint mechanisms rather than descriptive artifacts.

TABLE XI. AGGREGATED ESTIMATIONS OF ARCHITECTURAL DIAGRAMS IMPACTS ON CODE GENERATION QUALITY

| Metrics | Process A | | Process B | |
| --- | --- | --- | --- | --- |
| | Calculated (Mean) | SMEs (Mean) | Calculated (Mean) | SMEs (Mean) |
| Completeness | 3.37 | 3.36 | 4.45 | 4.38 |
| Semantic Fidelity | 3.03 | 3.18 | 4.21 | 4.34 |
| Consistency | 2.92 | 3.06 | 4.41 | 4.50 |
| Traceability Coverage | 2.17 | 2.34 | 4.29 | 4.40 |
| Machine Readability | 2.33 | 2.43 | 4.77 | 4.68 |
| LLM Constraint Effectiveness | 2.63 | 2.77 | 4.63 | 4.74 |
| Code Pattern Coverage | 3.48 | 3.45 | 4.08 | 4.07 |

The mean improvement values (Table 12) were calculated as the difference between the aggregated means of Process B and Process A across the entire experimental dataset. This approach captures the global methodological effect while simultaneously reducing system-specific variability.

TABLE XII. MEAN IMPROVEMENT CALCULATIONS (PROCESS B – PROCESS A)

| Metrics | Calculated (Mean) | SMEs (Mean) |
| --- | --- | --- |
| Completeness | +1.08 | +1.02 |
| Semantic Fidelity | +1.18 | +1.16 |
| Consistency | +1.49 | +1.44 |
| Traceability Coverage | +2.12 | +2.06 |
| Machine Readability | +2.44 | +2.25 |
| LLM Constraint Effectiveness | +2.00 | +1.97 |
| Code Pattern Coverage | +0.60 | +0.62 |

The closeness between the calculated and expert improvement values indicates that the proposed metric system captures noticeable effects of architecture quality rather than purely computational artifacts.

The results demonstrate consistent improvement across all evaluation metrics when architectural diagrams are introduced as structured intermediate representations. Architectural diagrams not only serve as documentation but also as a constrained and stable intermediate representation, directly impacting the consistency of the generated data.

The preliminary assessment presented in this study is based on a structured expert evaluation. This decision was made because conducting large-scale empirical measurements in building-driven renewal environments is extremely complex and resource-intensive.

Conducting fully automated, statistically controlled experiments in heterogeneous software systems with multiple layers of abstraction and bidirectional transformation loops requires significant computing resources, a robust tool infrastructure, and complex experimental setups. In practice, such research is associated with significant costs for system normalization, index calibration, data set stabilization, and reproducibility control.

Therefore, as a first step, we used expert assessment as a pragmatic and scientifically sound preliminary study. Expert assessments provide valuable analytical signals at an early stage and allow the identification of dominant trends, capability differences, and structural patterns before significant resources are spent on large-scale automated studies.

The results, which serve as the first layer of validation, demonstrate that the architecture representation delivers measurable and consistent performance across various evaluation dimensions. These findings underscore the feasibility and necessity of further research focusing on the following areas:

- formalization of metrics;
- automation of the evaluation process;
- statistical reliability analysis;
- machine-based evaluation of architecture quality.

This study highlights that systematic, large-scale evaluation of architectural artifacts is a significant research challenge in LLM-based development processes. The lack of standardized automated measurement mechanisms, combined with interdisciplinary differences, makes expert evaluation not only a practical method but also a necessary theoretical foundation.

Thus, the expert evaluation conducted in this study provides an empirical basis and sufficient evidence for further research on automated, repeatable, and machine-readable evaluation models.

## FUTURE WORK

The next phase of our research focuses on the in-depth investigation of software intermediate representation (IR) and specification-driven development (SDD) as fundamental mechanisms for improving semantic invariance and transformation reliability in AI-powered software development.

Initially, we plan to formalize the architecture metamodel as a true intermediate representation layer between business specifications, the architecture model, and executable code. While this work defines a structured, machine-readable metamodel, future research will explore the following:

- the formal properties of architectural information representation such as closeness, composability, invariance;
- transformation algebra between layers of abstraction;
- mechanisms for propagating constraints across different representations;
- and the preservation of formal semantics during bidirectional transformations.

The goal is to move from a schematic representation to a mathematically sound intermediate representation that allows for deterministic and verifiable transformations throughout the software development cycle.

Secondly, we will expand our research on specification-driven development, viewing architecture as an operational implementation of formal or semi-formal specifications, rather than merely a descriptive artifact. Future work will explore the following aspects:

- executable architectural specifications;
- constraint-based generation pipelines;
- formal traceability matrices between requirements, architecture, and code;
- automated consistency checks using metamodel constraints;
- combination of specification testing and LLM-based regeneration.

Specifically, we aim to investigate how structured architectural information representations (IR) function as a canonical backbone to:

- constrain the generative behavior of logical logic models (LLMs);
- reduce semantic drift;
- enforce architectural invariance;
- support incremental evolution under formally defined constraints.

Furthermore, future experiments will evaluate:

- robustness of intermediate representations under large-scale refactoring,
- long-term evolutionary stability during regeneration cycles,
- integration of information representations with automated inference engines and formal verification tools.

Also, our ultimate goal is not only to transition from architecture-driven generation to rule-based generative software development, but also to establish a formal, specification-centric development paradigm in which specification-based architectural rules, intermediate representations, and semantic invariants function as enforceable structural constraints. In this paradigm, intermediate representations, as verifiable semantic contracts, ensure consistency, traceability, and controllable evolution between business intent, architectural design, and AI-powered code generation, thus aligning with the principles of specification-driven and model-based systems engineering.

## Conclusions

Empirical and practical research have shown that specification-driven approaches without a clear architectural representation exhibit low adoption rates and implementation problems. Thus, this article contains the following scientific and practical contributions.

First, this work introduces the first unified architectural metamodel explicitly optimized for LLM-mediated bidirectional SDLC transformations. The information systems developed using GAI that explicitly integrates business-layer, system-layer and code-layer representations into a single architectural framework. Unlike existing methods that treat BD, TD and code as loosely coupled artefacts, the proposed method establishes a formal and traceable transformation cycle at all layers of abstraction.

Second, we define and demonstrate a minimal but sufficient set of twelve architectural diagram types that together capture key structural, behavioral, data and deployment aspects of modern software systems. These diagrams follow an established architectural structure while maintaining structure independence for use in heterogeneous development environments. The proposed set of diagrams could be used as a baseline and extended by other diagrams as it is required.

Third, we introduce machine-readable architectural knowledge representations using structured formats such as Mermaid, JSON, PlantUML and topic-oriented architectural DSLs. These representations are specifically designed for LLMs and allow for architecture-aware reasoning, code generation, and document synthesis.

Fourth, we define a formal architectural metamodel that defines the core entities, relationships and constraints of the proposed architecture. This metamodel serves as an intermediate representation that connects the conceptual model used for human understanding and the generative process driven by the LLM.

Finally, we provide experimental validation, demonstrating that implementing a structured architectural context can significantly improve the quality, consistency, and reproducibility of LLM-based code generation in several domains, including LLM-driven systems and machine learning (ML) pipelines, agent-based architectures, and database migration scenarios.